\documentclass[draftclsnofoot,onecolumn,12pt]{IEEEtran}
\usepackage{lipsum}

\usepackage[tbtags]{amsmath}
\usepackage{amsmath}

\usepackage[linesnumbered,ruled,commentsnumbered,longend]{algorithm2e}

\usepackage{mathtools}
\DeclarePairedDelimiterX{\inp}[2]{\langle}{\rangle}{#1, #2}

\usepackage{amssymb}
\usepackage{cite}
\usepackage{stfloats}

\usepackage{amsthm}
\usepackage{float}
\usepackage{amsthm}
\usepackage{amssymb}

\usepackage{graphicx}%
\usepackage{acronym}
\usepackage{balance}
\usepackage{stfloats}
\usepackage{color}
\usepackage{subcaption}

\usepackage{multirow}

\newcommand{\ub}[1]{\underbrace{#1}}

\newtheoremstyle{myremark}%
  {}
  {}
  {}
  {\parindent}
  {\itshape}
  {:}
  {5pt plus 1pt minus 1pt}
  {\thmname{#1}\thmnumber{~#2}\thmnote{~(#3)}}

\theoremstyle{myremark}

\theoremstyle{myremark}

\acrodef{CRN}{cognitive radio network}
\acrodef{SU}{secondary user}
\acrodef{PU}{primary user}
\acrodef{ZF}{zero forcing}
\acrodef{FD}{full-duplex}
\acrodef{BS}{base station}
\acrodef{i.i.d.}{independent and identically distributed}
\acrodef{DL}{downlink}
\acrodef{UL}{uplink}
\acrodef{SINR}{signal-to-interference-plus-noise ratio}
\acrodef{SNR}{signal noise ratio}
\acrodef{AWGN}{additive white Gaussian noise}
\acrodef{MMSE}{minimum mean square error}
\acrodef{SIC}{successive interference cancellation}
\acrodef{SI}{self-interference}
\acrodef{CCI}{co-channel interference}
\acrodef{MUI}{multiuser interference}
\acrodef{NOMA}{non-orthogonal multiple access}
\acrodef{OMA}{orthogonal multiple access}
\acrodef{QoS}{quality-of-service}
\acrodef{SIC}{successive interference cancellation}
\acrodef{SVD}{singular value decomposition}
\acrodef{MIMO}{multiple-input multiple-output}
\acrodef{SISO}{single-input single-output}
\acrodef{MIMO-NOMA}{multiple-input multiple-output non-orthogonal multiple access}
\acrodef{MIMO-OMA}{multiple-input multiple-output orthogonal multiple access}
\acrodef{PA}{power allocation}

\begin{document}

\title{Energy-Efficient Power Allocation in  Uplink mmWave Massive MIMO with NOMA}

\author{\IEEEauthorblockN{Ming Zeng,
Wanming Hao, Octavia A. Dobre, \emph{Senior Member}, \emph{IEEE}, and H. Vincent Poor, \emph{Fellow, IEEE} 
}
}

\maketitle

\begin{abstract}
In this paper, we study the energy efficiency (EE) maximization problem for an uplink millimeter wave massive {\color{black}multiple-input multiple-output} system with non-orthogonal multiple access (NOMA). Multiple two-user clusters are {\color{black}formed} according to their channel correlation and gain difference, {\color{black}and NOMA is applied within each cluster}. Then, a hybrid analog-digital beamforming scheme is designed to lower the number of radio frequency chains at the base station (BS). On this basis, 
we formulate a power allocation (PA) problem to maximize the EE under users' quality of service requirements. 
An iterative algorithm is proposed to obtain the PA. Moreover, an enhanced NOMA scheme is {\color{black}also} proposed, by exploiting the global information at the {\color{black}BS}. Numerical results show that the proposed NOMA schemes achieve superior EE when compared with {\color{black}the conventional orthogonal multiple access} scheme.
\end{abstract}


{\let\thefootnote\relax\footnote{
Copyright (c) 2015 IEEE. Personal use of this material is permitted. However, permission to use this material for any other purposes must be obtained from the IEEE by sending a request to pubs-permissions@ieee.org.

This work was supported in part by the Natural Sciences and Engineering Research Council of Canada (NSERC), though its Discovery program.

M. Zeng and O. A. Dobre are with Memorial University, St. John's, NL A1B 3X9, Canada (e-mail: mzeng, odobre@mun.ca).

W. Hao is with the School of Information Engineering, Zhengzhou
University, Zhengzhou 450001, China (e-mail: wmhao@hotmail.com).

H. V. Poor is with Princeton University, Princeton, NJ 08544 USA (e-mail: poor@princeton.edu)}  
}

\IEEEpeerreviewmaketitle

\section{Introduction}
Recently, millimeter wave (mmWave) communication has been recognized as a promising candidate for the fifth generation (5G) cellular networks \cite{Mumtaz}. {\color{black}Compared with the sub-6 GHz band communication, mmWave communication can provide orders-of-magnitude lager bandwidths by operating at 30-300 GHz, and thus}, {\color{black}can cope with} the explosive capacity demand for 5G. Except for the huge bandwidth, the smaller wavelengths at mmWave enable more
antennas to be packed in the same physical space, {\color{black}and thus, can better} support massive multiple-input multiple-output (mMIMO). This further introduces {\color{black}spatial} multiplexing and diversity gains. Indeed, it has been shown that mmWave mMIMO can attain orders-of-magnitude system capacity increment \cite{XGao}.  


Nonetheless, {\color{black}realizing mmWave mMIMO in practice still faces challenges \cite{Heath}: (1) to fully reap the gain provided by MIMO}, each antenna requires a dedicated radio-frequency (RF) chain, which is difficult to realize for mmWave due to space limitation. {\color{black}Furthermore}, the use of the massive number of antennas results in an equivalent number of RF chains, which is too expensive; (2) the power consumption of the RF chains can be unbearable, {\color{black}accounting for up to} $70\%$ of the total transceiver energy consumption \cite{Amadori}. 

To lower the transceiver complexity and energy consumption of mmWave MIMO, two mmWave-specific MIMO architectures have been proposed, {\color{black}namely} analog beamforming and hybrid analog-digital beamforming \cite{Wang_JSAC}. Compared with digital beamforming, these two architectures can significantly reduce the number of RF chains by performing the signal processing in analog or a mixture of analog-digital domains \cite{Heath}. Specifically, analog beamforming is often implemented using a network of phase shifters, and is one of the simplest approaches for applying MIMO in mmWave systems. However, its performance is compromised by the use of quantized phase shifts and the lack of amplitude adjustment. Hybrid analog-digital beamforming, on the other hand, can achieve a good balance between system complexity and performance, by appropriately selecting the number of RF chains. There are two main hybrid analog-digital beamforming schemes: one is the fully connected structure, {\color{black}which} connects all antennas to each RF chain, while the other is the partially connected scheme, {\color{black}which} divides the antennas into subarrays and only connects one subarray to its own RF chain \cite{Heath, Hao_access}.


However, as the number of RF chains decreases, the maximum number {\color{black}of users} that can be supported by mmWave mMIMO decreases as well. To break this fundamental limit and increase the number of simultaneously supported users, non-orthogonal multiple access (NOMA) has been integrated into mmWave mMIMO recently \cite{RA, 23, 3, 26, Zhu_uplink, 22}. Different from the conventional orthogonal multiple access (OMA), NOMA allows multiple users to access the same time-frequency resource by applying superposition coding and successive interference cancellation (SIC) \cite{18, 25, EE}. 
In addition, the use of NOMA in mmWave is preferable because the users' channels can be highly correlated due to the highly directional feature of mmWave transmission. \cite{26} considers downlink mmWave MIMO with hybrid beamforming, while \cite{Zhu_uplink} considers uplink with analog beamforming. Both {\color{black}works} focus on the spectral efficiency (SE) maximization problem. In contrast, \cite{22} investigates the energy efficiency (EE) maximization problem for a downlink mmWave MIMO with hybrid beamforming.

Different from all these previous works, we consider the EE maximization for an uplink NOMA-assisted mmWave mMIMO system. Compared with downlink, EE in uplink is of higher priority as user terminals are power-constrained. Multiple two-user clusters are formed according to their channel correlation and gain difference, and NOMA is applied within each cluster. Following this, we propose a hybrid analog-digital beamforming scheme to reduce the number of RF chains at the base station (BS). On this basis, a power allocation (PA) problem aiming to maximize the EE under users' quality-of-service (QoS) requirements is formulated. An iterative algorithm is proposed, which consists of an outer and inner loop. Moreover, considering that the BS has global information, we propose to further remove the interference across clusters. Then, the PA problem can still be solved by the proposed iterative algorithm. Simulation results show that {\color{black}our algorithms} achieve superior EE performance when compared with the conventional OMA scheme. 


%
%

\begin{figure}[t]
\begin{center}
\includegraphics[width=8cm,height=3cm]{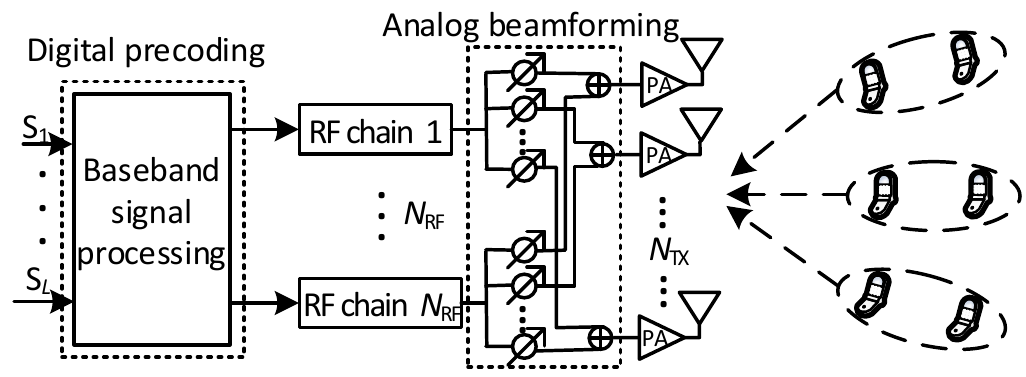}
\caption{Cluster-based uplink mmWave mMIMO-NOMA system model.}
\label{system}
\end{center}
\end{figure}

\section{System and Channel Model}
As shown in Fig. \ref{system}, we consider an uplink mmWave mMIMO-NOMA transmission scenario, where a BS communicates with $L$ two user clusters. 
The BS is equipped with $N_{\rm{TX}}$ antennas and $N_{\rm{RF}}$ RF chains satisfying ${N_{\rm{RF}}} \leq  N_{\rm{TX}}$, while all users are equipped with single antennas. It is assumed that the number of clusters equals that of RF chains ($L\!=\!N_{\rm{RF}}$). On this basis, 
users within the same cluster will be supported by the same beamforming vector. The full channel state information (CSI) is assumed available at the BS~\cite{XGao, O_Ayach}.

Owing to the limited scattering in mmWave channel, we adopt the geometric channel model with $F$ scatterers, where each scatter is assumed to contribute to a single propagation path between the user and the BS~\cite{XGao, O_Ayach}. Accordingly, the channel {\color{black}between the BS and the $i$-th user in the $l$-th cluster}, denoted as ${\bf{h}}_{l,i}, l\!\in\!\{1,\!\cdots\!,L\}$, $i\!\in\!\{1,2\}$, can be expressed as
\begin{equation}
{\bf{h}}_{l,i}\!=\!\sqrt{{N_{\rm{TX}}}/{F}}\sum_{f=1}^F\beta_{l,i}^f {\bf{a}}(\theta_{l,i}^f),
\end{equation}
where $\beta_{l,i}^f$ represents the complex gain of the $f$-th path, which is assumed to follow the Rayleigh distribution with zero mean and variance $\sigma_{f}$; $\theta_{l,i}^f\in[0,2\pi]$ is the $f$-th path's azimuth angle of arrival (AoA), while ${\bf{a}}(\theta_{l,i}^f)$ is the corresponding antenna array steering vector. We only consider the azimuth here, but the extension to elevation and azimuth is possible \cite{Heath}. For a uniform linear array configuration, ${\bf{a}}(\theta_{l,i}^f)$ is given by
${\bf{a}}(\theta_{l,i}^f)={{1}/{\sqrt{N_{\rm{TX}}}}}[1,e^{j\frac{2\pi}{\lambda}d\sin(\theta_{l,i}^f)},\ldots, e^{j\frac{2\pi}{\lambda}(N_{\rm{TX}}-1)d\sin(\theta_{l,i}^f)}]^T$,
where $\lambda$ and $d$ represent the signal wavelength and inter-antenna spacing, respectively \cite{4}. $[\cdot]^T$ denotes the transpose operation.

\section{Proposed Hybrid Beamforming Scheme}
{\color{black}
The received signals for the $l$-th cluster at the BS can be expressed as
\begin{align}\label{eq received singal}
{{y_{l}}}   &=\sum_{j=1}^L \sum_{i=1}^2 {\bf{v}}_l {\bf{B}}  {\bf{h}}_{j,i}  \sqrt{P_{j,i}} x_{j,i} + n_{l}  \\
&=  \sum_{i=1}^2 {\bf{v}}_l {\bf{B}}  {\bf{h}}_{l,i}   \sqrt{P_{l,i}} x_{l,i} + \sum_{j \neq l} \sum_{i=1}^2 {\bf{v}}_l {\bf{B}}  {\bf{h}}_{j,i}  \sqrt{P_{j,i}} x_{j,i} + n_{l}, \nonumber
\end{align}
where $x_{j,i}$ denotes the transmitted signal of user $i$ for the \text{$j$-th} cluster, satisfying $ \mathbb{E}(|x_{j,i}|^2)=1$. $P_{j,i}$ is the corresponding transmit power. ${\bf{B}}\in \mathbb{C}^{N_{\rm{RF}} \times  N_{\rm{TX}} }$ denotes the analog beamforming matrix for all clusters. ${\bf{v}}_l \in  \mathbb{C}^{1 \times N_{\rm{RF}}}$ is the digital beamforming vector for the $l$-th cluster. $n_{l}$ is the \ac{i.i.d.} additive white complex Gaussian noise (AWGN) with zero mean and variance $\sigma^2$. Additionally, the second term represents the inter-cluster
interference. Without loss of generality, it is assumed that $\| {\bf{v}}_{l} {\bf{B}} {\bf{h}}_{l,1} \|  \geq \| {\bf{v}}_{l} {\bf{B}} {\bf{h}}_{l,2} \|$, i.e., user 1 and user 2 represent the strong and weak users for the $l$-th cluster,~respectively. $\| \cdot \|$ denotes the Frobenius norm. }

Unlike the conventional analog beamforming design for OMA~\cite{O_Ayach}, where one analog beamforming vector is designed only for one user, two users need to share one analog beamforming vector in NOMA. Because of this, we should select a beamforming vector from the codebook of the considered two users that matches best their overall channels. Since the codebook should have the same form as the array steering vector ${\bf{a}}(\theta_{l,i}^f)$, we define the codebook of the $l$-th cluster as $\mathcal{F}_l=\{{\bf{a}}(\theta_{l,i}^f),f\in\{1,\!\cdots\!,F\},i\in\{1,2\}\}$. Based on this, the analog beamforming vector of the $l$-th cluster is selected in accordance to the following criterion:
\begin{eqnarray}\label{eq5}
{\bf{f}}_{{\rm{RF}},l}^\star=\underset{{\bf{f}}_{{\rm{RF}},l}\in\mathcal{F}_l}{\rm{arg\;max}} \;\;\;\| {\bf{f}}_{{\rm{RF}},l}^T {\bf{h}}_{l,1} \|
+\| {\bf{f}}_{{\rm{RF}},l}^T {\bf{h}}_{l,2}\|.
\end{eqnarray}

Accordingly, the overall analog beamforming matrix is given by ${\bf{B}}=[{\bf{f}}_{{\rm{RF}},1}^\star,\cdots,{\bf{f}}_{{\rm{RF}},L}^\star]^T$.
On this basis, we can obtain the equivalent uplink channel of each user as $\widetilde{{\bf{h}}}_{l,i}\!=\!{\bf{B}} {\bf{h}}_{l,i}$. With two users inside each cluster, the inter-cluster interference cannot be completely cancelled by the digital beamforming ~\cite{B_kim}. As in~\cite{B_kim}, to perform SIC correctly, we design the digital beamforming considering only the channels of the strong users, namely ${\bf{H}}=[\widetilde{{\bf{h}}}_{1,1},\!\cdots\!,\widetilde{{\bf{h}}}_{L,1}]$. Specifically, we generate the zero-forcing beamforming matrix ${\bf{V}}={\bf{H}}^H({\bf{H}}{\bf{H}}^H)^{-1}$, and apply the beamforming vector ${\bf{v}}_l={{\bf{V}}(l)}/{\|{\bf{V}}(l) {\bf{B}}\|}$ to the $l$-th cluster, where ${{\bf{V}}(l)}$ is the $l$-th row of ${{\bf{V}}}$. As a result, users in the $l$-th cluster still receive interference from the weak users in all the other clusters.

%

\section{Problem Formulation And Proposed Solution}
\subsection{Problem Formulation}
After {\color{black}hybrid} beamforming, the received signal-to-interference-plus-noise ratio (SINR) of users $i=1, 2$ in the $l$-th cluster can be written as
{\color{black}
\begin{eqnarray}\label{eq3}
{\rm{SINR}}_{l,1}=\frac{\| {\bf{v}}_l{\bf{B}} {\bf{h}}_{l,1}\|^2P_{l,1}}{\|{\bf{v}}_l{\bf{B}} {\bf{h}}_{l,2} \|^2P_{l,2}+\sum\limits_{j\neq l}\| {\bf{v}}_l{\bf{B}} {\bf{h}}_{j,2} \|^2P_{j,2}+\sigma^2},
\end{eqnarray}

\begin{eqnarray}\label{eq32}
{\rm{SINR}}_{l,2}=\frac{\| {\bf{v}}_l{\bf{B}} {\bf{h}}_{l,2} \|^2P_{l,2}}{\sum_{j\neq l}\| {\bf{v}}_l{\bf{B}} {\bf{h}}_{j,2} \|^2P_{j,2}+\sigma^2}.
\end{eqnarray}
}

{\color{black}
Denote $\rho_{l}=\| {\bf{v}}_l{\bf{B}} {\bf{h}}_{l,1} \|^2/{\sigma^2}$ and $\alpha_{j,l}=\|{\bf{v}}_l{\bf{B}} {\bf{h}}_{j,2} \|^2/\sigma^2$. 
Accordingly, their achievable rates are given by
\begin{equation}
R_{l,1}=\log_2 \left(1\!+\frac{\rho_{l} P_{l,1}}{\alpha_{l,l} P_{l,2}+\sum_{j\neq l}\alpha_{j,l}P_{j,2}+1} \right) ,
\end{equation}
\begin{equation}
R_{l,2}=\log_2 \left(1\!+\frac{\alpha_{l,l} P_{l,2}}{\sum_{j\neq l}\alpha_{j,l}P_{j,2}+1} \right) .
\end{equation}
}

{The total power consumption includes two parts: the flexible transmit power $\sum_{l=1}^L(P_{l,1}+P_{l,2})$, and the fixed circuit power consumption $P_{\rm{C}} $ \cite{Heath}.} 
The EE of the system is defined as
\begin{eqnarray}\label{eq9}
\eta_{\rm{EE}}=\frac{\sum_{l=1}^L(R_{l,1}+R_{l,2})}{\psi\sum_{l=1}^L(P_{l,1}+P_{l,2})+P_{\rm{C}}},
\end{eqnarray}
where $\psi$ is a constant accounting for the inefficiency of the power amplifier~\cite{K_ng}.
Finally, the EE maximization problem is formulated as follows:
{\color{black}
\begin{subequations}\label{eq10}
\begin{align}
&\underset{\bf{P}}{\rm{max}}\;\;\;\;\eta_{\rm{EE}}\\
{\rm{s.t.}}\;\;&R_{l,i}\geq R_{\rm{min}}, i\in\{1,2\}, l\in\{1,\cdots,L\},\label{eq101}\\
\;\;&P_{l,i} \leq P_{\rm{max}},i\in\{1,2\}, l\in\{1,\cdots,L\}\label{eq102},
\end{align}
\end{subequations}
where {\color{black}${\bf{P}} \in \mathbb{R}^{L \times 2}$} is the matrix of $P_{l,i}$. (\ref{eq101}) reflects {\color{black}the} QoS requirement for each user, while (\ref{eq102}) restricts the transmit power for each user to a maximum transmit power.}

\subsection{Proposed Solution}
{\color{black}It is clear that (\ref{eq10}) belongs to a fractional problem, which can be transformed into a series of parametric subtractive-form subproblems as follows:
\begin{align} \label{eq11}
\underset{\bf{P}} {\rm{max}}~ &{\sum_{l=1}^L(R_{l,1}\!+\!R_{l,2})}\!-\!\lambda^{(k-1)}\left({\psi\sum_{l=1}^L(P_{l,1}\!+\!P_{l,2})\!+\!P_{\rm{C}}}\right) \nonumber \\
{\rm{s.t.}} &\;(\rm{\ref{eq101}}),(\rm{\ref{eq102}}),
\end{align}
where $\lambda^{(k-1)}$ is a non-negative parameter. Starting from $\lambda^{(0)}=0$, $\lambda^{(k)}$ can be updated by $\lambda^{(k)} = \frac{\sum_{l=1}^L(R_{l,1}^{(k)}+R_{l,2}^{(k)})}{\psi\sum_{l=1}^L(P_{l,1}^{(k)}+P_{l,2}^{(k)})+P_{\rm{C}}}$, where $R_{l,i}^{(k)}$ and $P_{l,i}^{(k)}$ are the updated rates and power after solving (\ref{eq11}). Moreover, the maximum value of (\ref{eq11}) is calculated as $\varepsilon^{(k)}\! = \!\sum_{l=1}^L(R_{l,1}^{(k)}+R_{l,2}^{(k)}) \!-\!\lambda^{(k-1)} (\psi\sum_{l=1}^L(P_{l,1}^{(k)}+P_{l,2}^{(k)})+P_{\rm{C}})$. As shown in \cite{W_dink}, $\lambda^{(k)}$ keeps growing while $\varepsilon^{(k)}$ keeps declining as $k$ increases. When $\varepsilon^{(k)}=0$, $\lambda^{(k)}$ is maximized, which is also the maximum EE of (\ref{eq10}).   
}

Then, the problem consists in how to solve (\ref{eq11}) for a given $\lambda$. It is clear that (\ref{eq11}) is a non-convex optimization problem due to the non-concave objective function. We equivalently transform  (\ref{eq11}) as follows:
\begin{equation} \label{eq13}
\underset{\bf{P}}{\rm{max}}\;\;\sum_{l=1}^L\left(R_{l,1}+R_{l,2}-\lambda\psi(P_{l,1}+P_{l,2})\right),\;{\rm{s.t.}}\;(\rm{\ref{eq101}}),(\rm{\ref{eq102}}).
\end{equation} 

{\color{black}Let us consider the two users in the $l$-th cluster. After some mathematical manipulations, their sum rate is given by
\begin{equation}
R_{l,1}+R_{l,2} =\log_2 \left(\frac{\rho_{l} P_{l,1}+ \sum_{j=1}^L \alpha_{j,l} P_{j,2}+1}{\sum_{j\neq l}\alpha_{j,l}P_{j,2}+1} \right).
\end{equation}

By substituting the above equation into the objective function, we obtain
\begin{align}
f=&\ub{ \sum_{l=1}^L \log_2 \left({\rho_{l} P_{l,1}+ \sum_{j=1}^L \alpha_{j,l} P_{j,2}} +1\right)  - \lambda\psi \sum_{l=1}^L \left( P_{l,1}+P_{l,2} \right)  }_{f_1(\mathbf{P})} \nonumber \\
&-\ub{ \sum_{l=1}^L  \log_2 \left({\sum_{j\neq l}\alpha_{j,l}P_{j,2}+1} \right) }_{f_2(\mathbf{P})}.
\end{align}

\begin{algorithm}[t]
\caption{{\small{Energy-Efficient Power Allocation Algorithm}}}
\label{algorithms3}
{\bf{Initialize}}  $\varepsilon \leftarrow 10^{-6}$, $\lambda \leftarrow 0$. \\
\Repeat($\left\{\rm{Outer\;iteration}\right\}$){$\varepsilon^\star\leq \varepsilon$}
{Initialize feasible power ${\bf{P}}^{(0)}$.\\
\Repeat($\left\{\rm{Inner\;iteration}\right\}$){${\bf{P}}^{(k)}$ {\rm{converges}}}
 {${\bf{P}}^{(k)} \leftarrow {\rm{max}} ~~f_1({\bf{P}})-f_2({\bf{P}}^{(k-1)})- \inp{\nabla f_2({\bf{P}}^{(k-1)})}{{\bf{P}} -{\bf{P}}^{(k-1)}}$~${\rm{s.t.}} \; \eqref{rate_power_b}, \eqref{rate_power_c}, \eqref{eq102}$}
Compute $\varepsilon^{\star}\! \leftarrow \!\sum_{l=1}^L(R_{l,1}\!+\!R_{l,2})\!-\!\lambda(\psi\sum_{l=1}^L(P_{l,1}\!+\!P_{l,2})\!+\!P_{\rm{C}})$.\\
Update $\lambda  \leftarrow \frac{\sum_{l=1}^L(R_{l,1}+R_{l,2})}{\psi\sum_{l=1}^L(P_{l,1}+P_{l,2})+P_{\rm{C}}}$.\\
}
\end{algorithm}




Now let us consider the QoS constraints (\rm{\ref{eq101}}). After some mathematical manipulations, they can be reformulated as
\begin{align} 
& \rho_{l} P_{l,1}- \left(2^{R_{\rm{min}}}-1 \right) \left[\sum_{j=1}^L \alpha_{j,l} P_{j,2} +1 \right] \geq 0  \label{rate_power_b}\\
& \alpha_{l,l} P_{l,2}- \left(2^{R_{\rm{min}}}-1 \right) \left[\sum_{j \neq l} \alpha_{j,l} P_{j,2} +1 \right] \geq 0 \label{rate_power_c}.
\end{align}

Rewrite \eqref{eq13} as
\begin{equation} \label{dc_programming}
\underset{\bf{P}}{\rm{max}}~~ f_1(\mathbf{P})-f_2(\mathbf{P}), ~{\rm{s.t.}}~ \eqref{rate_power_b}, \eqref{rate_power_c}, (\rm{\ref{eq102}}),
\end{equation}
where both functions $ f_1(\mathbf{P})$ and $f_2(\mathbf{P})$ are concave. Thus, the objective $f_1(\mathbf{P})-f_2(\mathbf{P})$ is a DC function (difference of two concave functions). For $l \in \{1, \cdots, L\}$, define the vector $\mathbf{e}_l \in \mathbb{R}^{L}$, satisfying $\mathbf{e}_l(l) = 0$ and $\mathbf{e}_l(j) = \frac{\alpha_{j,l}}{\ln2}, ~ j \neq l$. The gradient of $f_2$ at $\mathbf{P}$ is given by
\begin{equation}
\nabla f_2({\mathbf{P}})= \sum_{l=1}^L \frac{1}{1+ \sum_{j\neq l}\alpha_{j,l}P_{j,2}} \mathbf{e}_l.
\end{equation} 

The following procedure generates a sequence $\{{\bf{P}}^{(k)} \} $ of improved feasible solutions. Initialized from a feasible $\{{\bf{P}}^{(0)} \} $, $\{{\bf{P}}^{(k)} \} $ is obtained as the optimal
solution of the following convex problem at the $k$-th iteration:
\begin{align} \label{dc_convex}
\underset{\bf{P}} {\rm{max}} ~~ &f_1({\bf{P}})-f_2({\bf{P}}^{(k-1)})- \inp{\nabla f_2({\bf{P}}^{(k-1)})}{{\bf{P}} -{\bf{P}}^{(k-1)}} ~  \nonumber \\
{\rm{s.t.}} &\; \eqref{rate_power_b}, \eqref{rate_power_c}, \eqref{eq102},
\end{align}
{\color{black}where $ \inp{\cdot}{\cdot}$ denotes the inner product operation.} 
Note that \eqref{dc_convex} can be efficiently solved by available convex software packages \cite{Cvx}.



\subsection{Complexity and Convergence}
The proposed algorithm includes inner and outer iterations. For the inner iteration, i.e., the DC programming, 
its convergence has been shown in \cite{DC}. 
For the outer iteration, i.e., the fractional programming, it always converges to the stationary and optimal solution~\cite{W_dink}. Therefore, the proposed algorithm always converges. 
The specific procedure of the proposed two layer algorithm is summarized in Algorithm 1.

Now, {\color{black}we discuss its computational complexity}. Denote the number of inner and outer iterations as $I_1$ and
$I_2$, respectively. The overall computational complexity of the proposed
algorithm is $O(D^2 I_1 I_2)$, where $D$ is the number of
dual variables for solving \eqref{dc_convex}.


\section{Enhanced NOMA}
In uplink, since the BS has global information, when the weak user in one cluster is decoded, its interference to users in other clusters can be removed. By applying this, the SINR can be further enhanced. Nonetheless, how to select the appropriate decoding order to maximize the EE is combinatorial and non-trivial. Here, a greedy algorithm is proposed.

{\color{black}Specifically, we denote the interference from user $(l,2)$ to other clusters as $\Gamma_l= \sum_{j \neq l} \alpha_{l,j} $. Without loss of generality, we arrange the clusters based on the {descending} order of $ \Gamma_l $.
To relieve the interference, the cluster with smaller index should be decoded earlier. After removing the inter-cluster interference from the already decoded users}, the SINR of user $(l,1)$ can be re-expressed as
\begin{eqnarray}\label{increased sinr}
{\rm{SINR}}_{l,1}^{'}=\frac{\rho_{l} P_{l,1}}{\alpha_{l,l} P_{l,2} + \sum_{j=l+1}^L \alpha_{j,l} P_{j,2}+1},
\end{eqnarray}
where $ \sum_{j=l+1}^L \alpha_{j,l} P_{j,2}$ denotes the remaining interference. Likewise, the SINR of user $(l,2)$ can be re-expressed as
\begin{eqnarray}\label{increased sinr}
{\rm{SINR}}_{l,2}^{'}=\frac{\alpha_{l,l} P_{l,2}}{\sum_{j=l+1}^L \alpha_{j,l} P_{j,2}+1}.
\end{eqnarray}

After some mathematical manipulations, the sum rate for the $l$-th cluster is given by 
\begin{equation}
{R}_{l,1}^{'}+{R}_{l,2}^{'} =\log_2 \left(\frac{\rho_{l} P_{l,1}+ \sum_{j=l}^L \alpha_{j,l} P_{j,2}+1}{\sum_{j=l+1}\alpha_{j,l}P_{j,2}+1} \right).
\end{equation}

Likewise, the corresponding QoS constraints are given by
\begin{align} 
& \rho_{l} P_{l,1}- \left(2^{R_{\rm{min}}}-1 \right) \left[\sum_{j=l}^L \alpha_{j,l} P_{j,2} +1 \right] \geq 0  \label{2rate_power_b}\\
& \alpha_{l,l} P_{l,2}- \left(2^{R_{\rm{min}}}-1 \right) \left[\sum_{j=l+1}^L \alpha_{j,l} P_{j,2} +1 \right] \geq 0. \label{2rate_power_c}
\end{align}

It can be seen that the PA problem after enhanced SINR has the same form as before. Therefore, here we can directly apply the proposed PA algorithm to obtain the solution.
}

%

%

\begin{table} [!h]
\caption{Simulation Parameters.} 
\renewcommand{\arraystretch}{0.75}
\label{Table III} 
\centering
 \begin{tabular}{c|c} 
 \hline  
\bfseries Parameters & \bfseries Value \\ [0.5ex] 
 \hline\hline
 Number of antennas  &$N_{\rm{TX}}=100$ \\
 \hline
 Number of RF chains   &$N_{\rm{RF}}=8$ \\
 \hline
 Minimum rate requirement &$R^{\rm{min}}=0.1$  [bit/s/Hz] \\
 \hline
 Fixed transmit power per user & $P_f = 10$ [dBm] \\
 \hline
 Maximum transmit power per user & $P_{\rm{max}} = 10$ [dBm]  \\
 \hline
 Inefficiency factor & $\psi=1/0.38$ \\
 \hline  
 Channel bandwidth & $50$ [MHz] \\ 
 \hline
Thermal noise density & $-174$ [dBm/Hz] \\
 \hline
 Path-loss exponent & 4.3 \\
 \hline 
 Cell radius & $0.3$ [km]\\
 \hline  
\end{tabular}
\end{table}

\section{Simulation Results}
In this section, simulations are conducted to evaluate the effectiveness of the proposed mmWave mMIMO-NOMA schemes. The default simulation parameters are listed in \text{Table I}. The users are randomly placed within the cell radius following a uniform distribution, and the azimuth AOA is uniformly distributed over $[0,2\pi]$. In terms of user pairing, as in \cite{B_kim, 22}, we first define the channel {\color{black}correlation} and gain difference between users $i$ and $j$ as ${\rm{Corr}}_{(i,j)}=|{\bf{h}}_i{\bf{h}}_j^T|/\|{\bf{h}}_i\|\|{\bf{h}}_j\|$ and $\pi_{(i,j)}=\left|\|{\bf{h}}_i\|-\|{\bf{h}}_j\|\right|$, respectively. Then, we select the user pairs with larger channel correlation. When two user pairs have the same channel correlation, we select the pair with a larger channel gain difference. A total of $L=8$ two user pairs are selected.
As for baseline algorithms, we represent the conventional OMA through time-division multiple access, where equal time slots are allocated to users in the same cluster, and the maximum EE is obtained via solving the corresponding optimization problem. 

 


Fig. 2(a) shows how the EE varies with the maximum transmit power $P_{\rm{max}}$. 
{\color{black}``MaxEE" denotes the EE maximization results, while ``MaxSE" represents the obtained EE when the SE of the system is maximized, i.e., $\lambda=0$. ``E-NOMA'' denotes the enhanced NOMA scheme.} 
When $ P_{\rm{max}} \in [-10, -6]$ dBm, only E-NOMA can satisfy the QoS requirements. When $P_{\rm{max}} \in [-5, 0]$ dBm, {\color{black}the EE provided by} all three schemes grows with $P_{\rm{max}}$. When $P_{\rm{max}} \geq 0$ dBm, for all three schemes, the EE remains fixed for MaxEE, while it declines for MaxSE. This shows the necessity of employing energy-efficient PA, especially under high $P_{\rm{max}}$. In addition, it can be seen that the two NOMA schemes achieve much higher EE than OMA for all feasible $P_{\rm{max}}$ values. Moreover, for the two NOMA schemes, E-NOMA always outperforms NOMA.
Also, we can observe that there is a {\color{black}visible} gap between MaxEE and MaxSE for E-NOMA, but not for NOMA during the increasing phase. This is because in E-NOMA, clusters receive different levels of interference due to inter-cluster interference cancellation. This leads to some clusters having extra power left after maximizing the EE. However, for MaxSE, this extra power is also used for sum rate maximization, which yields a decreased EE compared with MaxEE. 

\begin{figure*}
\centering
\begin{subfigure}{0.5\textwidth}
  \centering
  \includegraphics[width=1\linewidth]{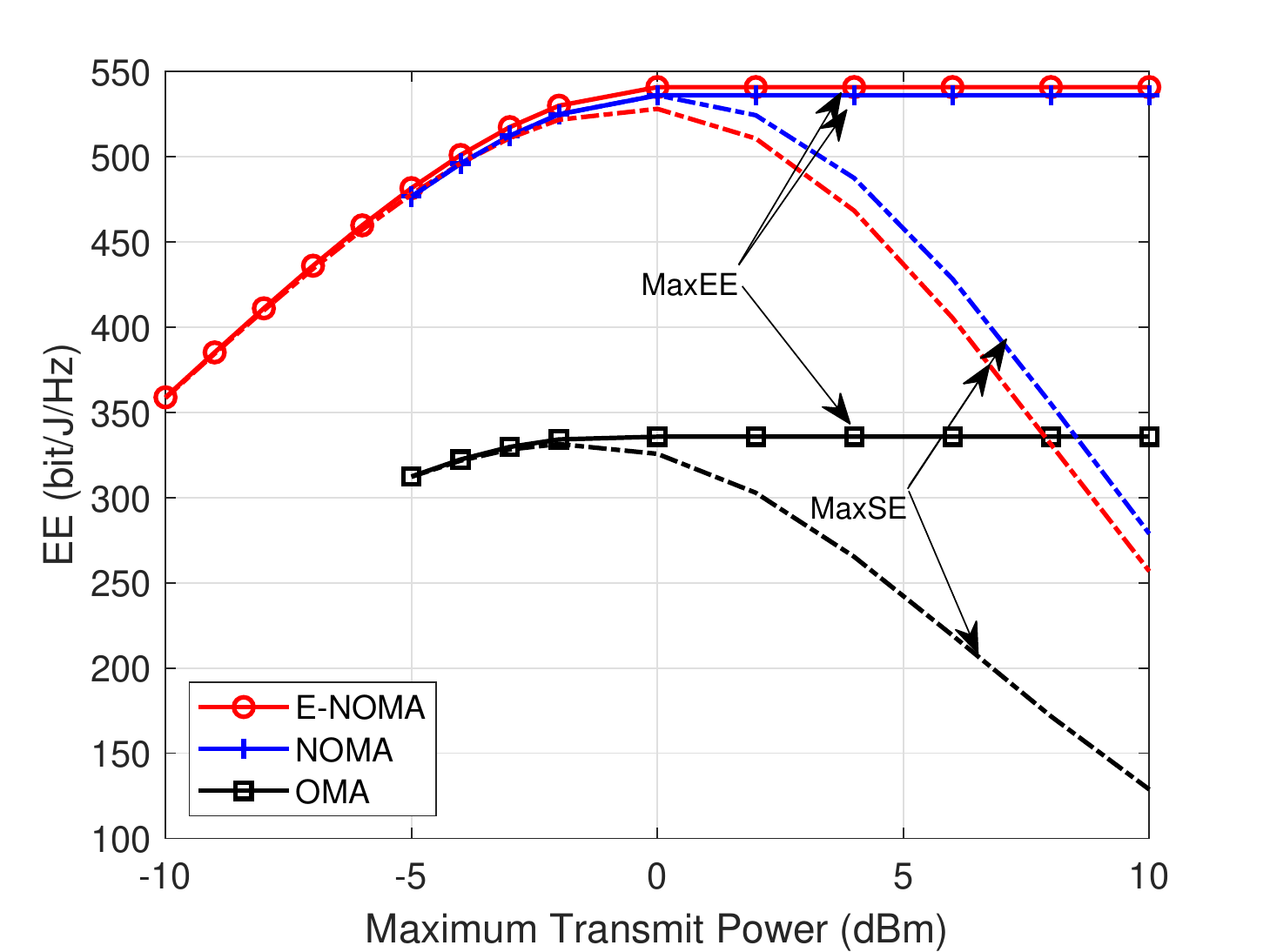}
  \caption{}
  \label{fig:sub1}
\end{subfigure}%
\begin{subfigure}{0.5\textwidth}
  \centering
  \includegraphics[width=1\linewidth]{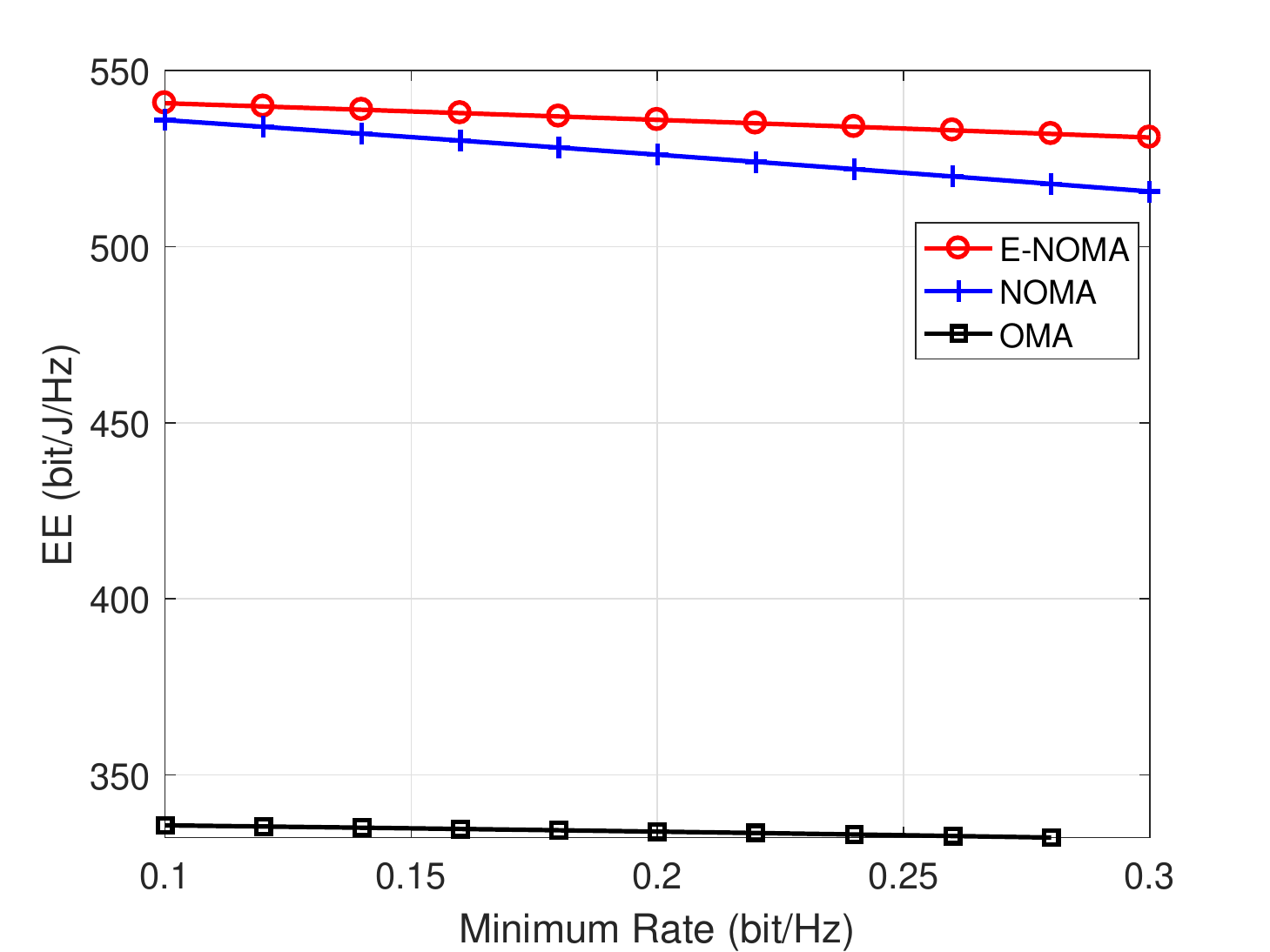}
  \caption{}
  \label{fig:sub2}
\end{subfigure}
\caption{EE comparison for NOMA and OMA: a) when the maximum transmit power varies; $R^{\rm{min}}=0.1$ bit/s/Hz; b) when the minimum rate varies; $P_{\rm{max}}=0$ dBm.}
\label{fig:test}
\end{figure*}

Fig. 2(b) plots how the EE varies with the minimum rate requirement $R^{\rm{min}} $. It is clear that the EE of all three algorithms decreases as $R^{\rm{min}} $ increases. Among the three algorithms, the two NOMA schemes still achieve much higher EE than OMA. In particular, when $ R^{\rm{min}} =0.3$ bps/Hz, OMA is infeasible while the two NOMA schemes can still satisfy the QoS requirements. Furthermore, it can be seen that E-NOMA outperforms NOMA for all $ R^{\rm{min}}$ values.

\section{Conclusion}
In this paper, we investigated the EE maximization problem for uplink mmWave mMIMO-NOMA. A hybrid analog-digital beamforming scheme was first proposed to lower the number of RF chains at the BS. Then, an iterative algorithm was introduced to allocate the power for EE maximization. Moreover, by further removing the inter-cluster interference at the BS, an enhanced NOMA scheme was presented, and the same algorithm can be applied for PA. Simulation results showed that both NOMA schemes achieve much higher EE than OMA as the maximum transmit power and minimum rate requirements vary. Furthermore,
the proposed E-NOMA always outperforms the conventional NOMA in terms of EE.

\bibliographystyle{IEEEtran}
\balance
\bibliography{IEEEabrv,conf_short,jour_short,mybibfile}

\end{document}